%% file: main-superres.tex
\documentclass[number,sort&compress,5p,authoryear,twocolumn]{elsarticle}

\usepackage[utf8]{inputenc}
\usepackage[T1]{fontenc}

\usepackage{amsmath}
\usepackage{amssymb}
\usepackage{amsfonts}
\usepackage{nicefrac}
\usepackage{mathtools}

\usepackage{tikz}
\usetikzlibrary{bayesnet}
\usetikzlibrary{arrows.meta}

\usepackage{graphicx}
\graphicspath{{./Figures/}}

\usepackage{subfig}

\usepackage[labelfont=bf]{caption}

\usepackage{textcomp}
\usepackage[euler]{textgreek}

\usepackage{bm}

\usepackage{nicefrac}
\usepackage{xcolor}
\usepackage{esvect}
\usepackage{widetable}
\usepackage{booktabs}
\usepackage{gensymb} 
\usepackage{dsfont}
\usepackage{algorithm,algpseudocode}
\usepackage[draft]{hyperref}

\newcommand\figref[1]{\hyperref[#1]{figure~\ref*{#1}}}
\newcommand\Figref[1]{\hyperref[#1]{Figure~\ref*{#1}}}
\makeatletter
\def\shorteqref{\@ifstar\@shorteqref\@@shorteqref}
\def\@shorteqref#1{\textup{\tagform@{\ref*{#1}}}}
\def\@@shorteqref#1{\hyperref[#1]{\textup{\tagform@{\ref*{#1}}}}}
\makeatother 
\renewcommand\eqref[1]{\hyperref[#1]{equation~\shorteqref*{#1}}}
\newcommand\Eqref[1]{\hyperref[#1]{Equation~\shorteqref*{#1}}}
\newcommand\tabref[1]{\hyperref[#1]{table~\ref*{#1}}}
\newcommand\Tabref[1]{\hyperref[#1]{Table~\ref*{#1}}}
\newcommand\secref[1]{\hyperref[#1]{section~\ref*{#1}}}
\newcommand\Secref[1]{\hyperref[#1]{Section~\ref*{#1}}}
\newcommand\supfigref[1]{\hyperref[#1]{inline supplementary 
figure~\ref*{#1}}}
\newcommand\Supfigref[1]{\hyperref[#1]{Inline supplementary 
figure~\ref*{#1}}}
\newcommand\suptabref[1]{\hyperref[#1]{inline supplementary 
table~\ref*{#1}}}
\newcommand\Suptabref[1]{\hyperref[#1]{Inline supplementary 
table~\ref*{#1}}}
\newcommand\supmatref[1]{\hyperref[#1]{inline supplementary 
material~\ref*{#1}}}
\newcommand\Supmatref[1]{\hyperref[#1]{Inline supplementary 
material~\ref*{#1}}}
\newcommand\apxref[1]{\hyperref[#1]{\ref*{#1}}}
\newcommand\Apxref[1]{\hyperref[#1]{\ref*{#1}}}

\let\originalcite\cite
\let\originalcitep\citep
\let\originalcitet\citet
\renewcommand\cite[1]{\nopagebreak\originalcite{#1}}
\renewcommand\citep[1]{\nopagebreak\originalcitep{#1}}
\renewcommand\citet[1]{\nopagebreak\originalcitet{#1}}

\DeclareMathOperator{\tr}{^{\mathrm{T}}}
\DeclareMathOperator*{\argmin}{\arg\!\min} 

\newcommand{\norm}[1]{\left\lVert#1\right\rVert} 
\newcommand{\vt}[1]{{\bf#1}} 
\newcommand{\vm}[1]{\bm{#1}} 
\newcommand{\inr}[1]{\in \mathbb{R}^{#1}} 
\newcommand{\N}[1]{\mathcal{N} \left( #1 \right)} 

\begin{document}

\input{front}

\section{Introduction}

Automated analysis of magnetic resonance imaging (MRI) data enables
studies of the brain that would be either very time consuming, or simply
not possible, if performed manually. These  automated methods can be
used to, \emph{e.g.}, investigate differences in tissue composition
between groups of subjects \citep{ashburner2000voxel}, examine changes
over time due to ageing and neurodegenerative disorders
\citep{reuter2012within}, make predictions between healthy and diseased
patients \citep{kloppel2008automatic} or delineate to a clinician the
location of some pathology \citep{kamnitsas2017efficient}. Many
neuroimaging software packages used for such analysis, \emph{e.g.}, SPM
\citep{ashburner2005unified}, FSL \citep{zhang2001segmentation} and
FreeSurfer \citep{fischl2004sequence}, are tailored for small, almost
isotropic voxels, with high contrast between grey matter (GM) and white
matter (WM). As they require more time and high field strengths, these
types of features are mostly found in images acquired in a controlled,
research context.

\begin{figure*}[t]
\centering
\includegraphics[width=\textwidth]{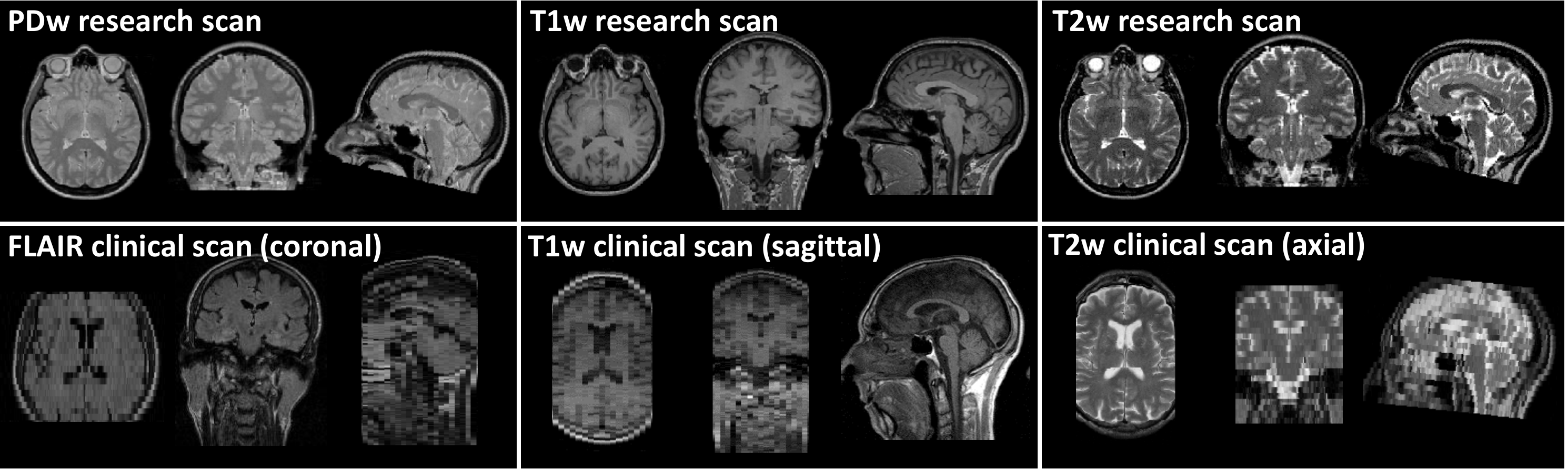}
\caption[Comparing research and clinical MRI]{Comparing research and
clinical MRI. The top row shows PD-weighted (PDw), T1-weighted (T1w) and
T2-weighted (T2w) MR images of a subject from the high-resolution IXI
dataset, which is used in the simulated results section of this paper.
The three scans have close to 1 mm isotropic voxels. The bottom row
shows FLAIR, T1w and T2w MR images of a subject from the clinical
dataset we use in our evaluation. These subject's scans have
thick-slices (6.5 mm), with different slice-select directions, and
partial brain coverage.}
\label{fig:superres:ex-res-vs-hosp}
\end{figure*}

MR images acquired in a clinical setting have much greater variability
than their research counterparts, as clinicians do not target
reproducibility, but aim at maximising sensitivity to a suspected
pathology. This variability can be disentangled into instrumental and
biological. The instrumental variability stems from: (1) clinicians
favouring speed over volume resolution and therefore typically acquiring
images with high in-plane resolution, but fewer and thicker slices; and
(2), the scans being acquired across a diverse set of MR sequences,
sensitive to different tissue properties and, therefore, different
pathologies. On the other hand, the biological variability relates to
the diverse morphological variability in patient brains due to different
clinical conditions, as well as a wide age distribution. All in all,
this leads to a huge diversity in clinical MRI data. A comparison
between MR images acquired in a research and a clinical context is shown
in Figure \ref{fig:superres:ex-res-vs-hosp}. Applying automated analysis
pipelines, successfully, to images acquired in a routine clinical
setting is much more challenging than applying them to research images.
For example, the aforementioned software packages were recently
evaluated at the task of brain GM and WM volume estimation from
isotropic and anisotropic volumes acquired in the same subjects, and
results were improved by several percentage points when isotropic
volumes were used over thick-sliced ones \citep{adduru2017leveraging}.

Conversely, researchers interested in applying automated analysis
methods to MR data trade speed for resolution and therefore tend to
acquire high-resolution (HR) volumetric images. This makes large-scale
studies, with thousands of subjects, difficult to perform due to the
expense of scanning with such parameters. For example, UK BioBank took a
decade of planning and acquisition \citep{ollier2005uk}, at a cost of
more than £60 million \citep{palmer2007uk}. However, such
studies can provide great opportunities for making new discoveries about
the brain \citep{van2013wu,miller2016multimodal,smith2018statistical}.
Clinical data does not have this problem, as huge amounts of
population-representative\footnote{Not the general population -- rather
the population who are likely to have had strokes, tumours, etc.} scans
exist \citep{roobottom2010radiation}, accumulated over years of clinical
service. Furthermore, these scans are available for research at cost
neutrality and have a much higher prevalence of pathology, which is of
interest if disease is to be investigated rather than normality.

Commonly, for analyses that require isotropic voxels, thick-sliced
volumes are upsampled using simple interpolation, even though it can
introduce artefacts such as aliasing \citep{aganj2012removing} and bias
subsequent analyses. In this paper we propose a method for
reconstructing HR images with isotropic voxels from multimodal,
low-resolution (LR) clinical scans, based on a principled generative
model. Carefully crafted generative models have been shown to generalise
well in the case of medical image segmentation
\citep{ashburner2005unified,zhang2001segmentation,fischl2004sequence},
and we here developed such a model for clinical MRI super-resolution.
The objective of our model is to merge information that is distributed
across all MR scans of a patient, in order to generate
closer-to-research quality images from clinical scans. Reconstructing
isotropic HR images in such a way could in turn better enable the
aforementioned large-scale studies. The multi-sequence capability of the
proposed model stems from a cross-channel\footnote{We consider the
different MR contrasts acquired in a patient as being equivalent to
colour channels in computer vision (\emph{e.g.}, RGB).} functional that
has been studied extensively in the computer vision community; in MRI,
for parallel image reconstruction \citep{chatnuntawech2016vectorial} and
recently introduced for multimodal super-resolution
\citep{brudfors2018mri}.

The model-based approach we propose in this paper is data agnostic and
does not require any complex learning. Learning a representative
distribution of the large number of sequences used in clinical MR
imaging is extremely challenging and would make the model much less
general. Having to train a model anew, each time a new combination of MR
contrasts is encountered, or a different scaling factor is needed, would
make a general tool much more difficult to develop.

We show on a large number of patient scans that the model is capable of
super-resolving very thick-sliced clinical MRIs ($\approx$6.5 mm), with
missing fields of view. To ensure good generalisability, all model
parameters were estimated in a principled way. Our implementation is
publicly available, and can be used by researchers interested in
analysing routine clinical MR data. The implementation is available from
\url{https://github.com/brudfors/spm_superres}, we intend to shortly
incorporate it into the SPM12
software\footnote{\url{www.fil.ion.ucl.ac.uk/spm/software/spm12}}.


\section{Super-Resolution in Brain MRI}

In image processing, the task of improving the resolution of an image
after it has been acquired is known as super-resolution\footnote{The
term super-resolution can also be used to describe techniques attempting
to transcend the diffraction limit of an optical system, \emph{e.g.},
super-resolution microscopy.}. There are two cardinal ways of
super-resolving: the first is to combine information from a multiplicity
of images of the same subject, the second is to introduce guidance from
a population of other subjects. For both these methods it common to use
some kind of inductive bias given mathematically or conceptually,
\emph{e.g.}, regularisation.

Super-resolution was first studied in the computer vision field in the
early 80s and has since grown to become an active area of research using
tools from, \emph{e.g.}, inverse problems and machine learning
\citep{huang2010image,yue2016image}. In the early 2000s,
super-resolution found applications in the medical imaging community, in
particular for brain MRI \citep{van2012super}. This was partly due to
multi-LR super-resolution's dependency on exact image alignment, where
for brain imaging it is possible to obtain good alignment simply by
rigid registration. It was furthermore shown that super-resolution
improved resolution and signal-to-noise ratio favourably, compared with
direct HR acquisition \citep{plenge2012super}.

The earliest works investigating super-resolution applied to brain MR
images, defined an observation model that did not take into account
multiple MR contrasts, but rather specified how a set of LR images, of
the same contrast, were generated from a HR image
\citep{peled2001superresolution,greenspan2001mri}. The basic idea was
that it should be more beneficial (with respect to the trade-off between
acquisition time and signal-to-noise ratio), to acquire multiple LR
images, and then combining them using super-resolution, than acquiring
one HR image. The observation model was constructed by linear operations
(shift, down-sampling, blurring) and additive Gaussian noise. A maximum
likelihood (ML) estimate of the unknown HR image could then be found by
a gradient descent style algorithm. Limited in handling scans only of
the same orientations, but with subpixel offsets,
\cite{shilling2009super} extended the methods based on observation
models to handle multiple LR images, of different slice-select
directions. The work of \cite{poot2010general} subsequently generalised
this approach to images not necessarily rotated around a common
frequency encoding axis, allowing for any slice orientation. They
additionally solved the optimisation problem efficiently using a
conjugate gradient algorithm.

The ML methods were quickly extended to include some form of
regularisation via maximum a posteriori (MAP) estimation, in order to
reduce the ill-posedness of the solution, as well incorporating prior
knowledge (\emph{e.g.}, neighbouring voxels should look similar). The
work of \cite{peeters2004use}, \cite{kainz2015fast} and
\cite{rousseau2010super} did so using edge-preserving functionals, such
as the Huber function or anisotropic diffusion, to regularise their
solutions, while \cite{zhang2008application} and \cite{poot2010general}
used the Euclidean norm of the image gradients. In \cite{bai2004super} a
Markov random field was used, and in \cite{shi2015lrtv} two sorts of
regularisation were combined: the edge-preserving total variation and a
low-rank term that enabled utilisation of information throughout the
image. Also \cite{tourbier2015efficient} used total variation as a 
regulariser, with an adaptive regularisation scheme. Besides 
including some 
sort of regularisation in the objective
function, replacing its mean-squared error norm has also been explored,
with the aim of increased robustness to incorrect noise models
\citep{gholipour2010robust}.

The methods discussed up until this point are non-optimal for processing
clinical MRIs, as they assume multiple LR images of the same contrast
and hence do not utilise the fact that routine clinical scans often are
of different contrasts. A method utilising information from a HR
reference image of a different contrast was first proposed by
\cite{rousseau2010non}. The super-resolved image was obtained by
iteratively denoising the current estimate of the super-resolved image
with a patch-based technique, using the HR image as a reference, and
then solving an optimisation problem, where the denoised image
regularised the solution. The patch-based work of \cite{manjon2010mri}
used the information from a HR reference in a similar way, but instead
of solving for the super-resolved image in an optimisation setting, they
used an iterative reconstruction-correction scheme. Another patch-based
approach estimated weights from a HR image of a given contrast and then
regressed a HR image from a LR image with a different contrast
\citep{zheng2017multi}. Numerous other works on super-resolving MRIs
have used these patch-based approaches, utilising the pattern redundancy
present in image patches
\citep{manjon2010non,coupe2013collaborative,plenge2013super,jog2014improving}.
 However, although these methods enable super-resolving across MR 
channels, they require access to HR data. Such data is seldom available 
in a clinical setting. To mitigate this problem, methods have been 
developed that utilises the property that clinical MR images are 
inherently anisotropic to learn a regression between LR and HR images 
\citep{jog2016self,zhao2018deep}; however, using only a single 
contrast.


Data-driven (or learning-based) methods have also been thoroughly
investigated for the task of super-resolving brain MRIs. These methods
aim at learning a LR to HR mapping from training data. One of the first
such methods was proposed by \cite{rueda2013single}, using sparse
dictionary learning. Also, regression-based techniques have been
explored, such as patch-based random forest regression
\citep{alexander2014image,sindel2018learning}. Convolutional neural
networks is another class of supervised methods that has gained interest
in MRI super-resolution
\citep{tanno2017bayesian,cengiz2017super,pham2017brain}. Deep generative
models have also been used to directly learn a LR to HR distribution for
super-resolution recovery \citep{chen2018efficient}. To combat the fact
that HR training data is difficult to obtain, \cite{dalca2018medical}
proposed a generative model for sparse image patches, where LR clinical
scans were used as training data. They additionally dealt with missing
data in an elegant way. However, to the best of our knowledge, there is
not yet a learning-based super-resolution tool useable without the need
for training anew, when faced with unseen contrasts.

\section{Methods}

Compared with some of the super-resolution approaches mentioned in the
previous section, the model we propose here it multimodal, and does not
require any complex learning, nor any HR reference data. We additionally
avoid any learning from training data with the hope of improved
generalisability when applying the model to diverse clinical MR scans.
The model instead relies on the definition of a joint distribution of a
patient's multimodal MR images. A generative model that integrates
multimodal images requires a component that relates signal across the
various modalities. Here, this component is a novel prior in the context
of MRI super-resolution, which promotes combining image information
distributed across a patient's MR images. In this section, we consider
that each MR contrast (\emph{e.g.}, T1w, T2w, PDw) constitutes one
channel of a multi-channel volume. When multiple images of the same
contrast are acquired, they are considered as multiple noisy
observations of the same channel.

\begin{figure}[t]
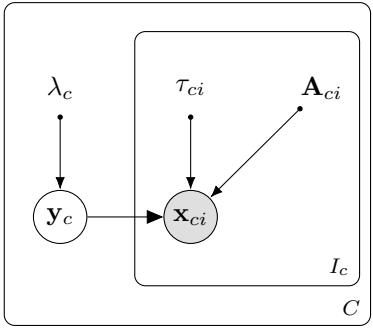

\centering
\tikz{
\node[latent] (y) {$\vt{y}_{c}$};
\node[latent, draw=white, above=of y] (lambda) {$\lambda_c$};
\node[obs, right=of y] (x) {$\vt{x}_{ci}$};
\node[latent, draw=white, above=of x] (tau) {$\tau_{ci}$};
\node[latent, draw=white, above=of x, right=of tau] (A) {$\vt{A}_{ci}$} ;

\plate[inner sep=0.25cm, xshift=-0.12cm, yshift=0.12cm] 
{plate1} {(x) (tau) (A)} {$I_c$};
\plate[inner sep=0.25cm, xshift=-0.12cm, yshift=0.12cm] 
{plate2} {(y) (lambda) (plate1)} {$C$}; 

\edge[{Circle[length=2pt]}-Latex] {lambda} {y};
\edge[{Circle[length=2pt]}-Latex] {tau} {x};
\edge[{Circle[length=2pt]}-Latex] {A} {x};
\edge {y} {x};
}
\caption[Graphical model of the generative model for super-resolving MR
images using a multi-channel prior]{Graphical model of the joint
probability distribution in \eqref{eq:superres:jp}. Random variables are
in circles, observed variables are shaded, plates indicate replication.
There are $C$ unknown HR images ($\vt{y}_c$) and $I_c$ 
observed LR images ($\vt{x}_{ci}$) for each $c$. Model parameters
are dotted and are the Gaussian noise precisions $\tau_{ci}$,
projection matrices $\vt{A}_{ci}$, and regularisation parameters
$\lambda_c$.}
\label{fig:superres:gm}
\end{figure}

\subsection{The Generative Model}

The model assumes that each LR image is generated by selecting thick
slices, arbitrarily rotated and/or translated, from a HR image, with the
addition of random noise. This assumption can be written as a
conditional probability distribution known as the data likelihood. We
also assume that each HR image is the result of a random process,
characterised by a probability distribution known as the prior. This 
generative model is formalised by the joint probability distribution:
\begin{align}
p(\vm{\mathcal{X}},\vt{Y}) =
\underbrace{p(\vm{\mathcal{X}} \mid 
\vt{Y})}_{\text{likelihood}}~\underbrace{p(\vt{Y})}_{\text{prior}} = 
\prod_{c=1}^C 
\prod_{i=1}^{I_c} p(\vt{x}_{ci} \mid \vt{y}_c)~p(\vt{Y}),
\label{eq:superres:jp}
\end{align}
where $\vt{Y}~=~\{ \{ \vt{y}_c \}_{c=1}^{C} \mid \vt{y}_c \in
\mathbb{R}^{N} \}$ denotes the unknown HR images of $C$ different
channels and $\vm{\mathcal{X}}~=~\{ \{ \vt{X}_c \}_{c=1}^{C} \mid
\vt{X}_c=\{ \vt{x}_{ci} \}_{i=1}^{I_c} \mid \vt{x}_{ci} \in
\mathbb{R}^{N_{ci}} \}$ a set of LR images. The variable $I_c$ is  the
number of observed LR images of channel $c$, $N$ is the number of voxels
in the HR images, and $N_{ci}$ is the number of voxels in the $i$th LR
image of the $c$th channel. Note that we allow for some values of
observed LR voxels to be assumed missing (\emph{i.e.}, Not a
Number), which enables these values to be filled in during model
fitting.  Prior to super-resolving a set of MRIs we perform a rigid
registration of the observed data to a common reference using the
\texttt{spm\_coreg} routine of SPM12\footnote{A more elegant way of
doing this would be to include registration inside the generative model,
such that fitting the model would optimise also some registration
parameters (see Discussion).}.

We cast estimating the HR images ($\vt{Y}$), given a set of observed LR
images ($\vm{\mathcal{X}}$), as MAP estimation in the joint probability
distribution defined by \eqref{eq:superres:jp}:
\begin{align}
&p({\bf Y} \mid \bm{\mathcal{X}})  = 
p(\bm{\mathcal{X}}\mid{\bf Y})~p({\bf Y})~/~p(\bm{\mathcal{X}}) \cr 
&\Rightarrow \quad \argmin_{\bf Y} \left\{ - \ln  p({\bf 
Y} \mid 
\bm{\mathcal{X}}) \right\} \cr
&\Rightarrow \quad \argmin_{\bf Y} \left\{ - \ln  
p(\bm{\mathcal{X}}\mid{\bf Y})~p({\bf Y}) \right\}.
\label{eq:superres:map}
\end{align}
For the model to generalise, we estimate its parameters from either the
observed data (likelihood and prior hyper-parameters) or set them in a
general and consistent way (projection matrices). A graphical
representation of the generative model is shown in Figure
\ref{fig:superres:gm}. As in practice, there is rarely more than one
observation of the same channel, we will from now on drop the summations
over $i$ and assume only one observation of each channel. All
derivations stay the same, except for additional summations over
conditional distributions of LR images. The individual components of our
generative model will be further explained in the next three sections.

\subsubsection{Model Likelihood}

The likelihood function should describe the data generating process of
LR images ($\vt{x}_{c}$) from unknown HR images. Its main component is a
deterministic projection matrix ($\vt{A}_c$) that encodes the
slice-selection parameters (orientation, thickness, gap, profile) of an
LR image. It is a linear operator that, when applied to the
corresponding HR image, creates a noiseless LR version. The second
component of the generative process encodes acquisition noise. As MR
images are usually reconstructed as the magnitude of an image that was
originally complex -- and Gaussian noise on complex data gives a Rice
distribution in the magnitude image -- a Rician noise model would be
suitable\footnote{Note that multi-coil MR images generally do not have
Rician noise because of the way the images are reconstructed.  The
Rician assumption is therefore good for older MR scanners, but becomes a
bit of an approximation for more modern systems.} \citep{aja2013review}.
However, it has been shown that the mathematically more tractable
Gaussian distribution closely approximates the true Rician noise
distribution in MRI \citep{gudbjartsson1995rician}. We therefore assume
the following forward model:
\begin{align}
\vt{x}_{c} &= \vt{A}_{c} \vt{y}_c + \epsilon, \qquad \epsilon \sim 
\N{0,\tau_{c}^{-1}}, 
\label{eq:superres:noisemodel}
\end{align}
so that the conditional distribution of an observed LR image, given an
unknown HR image, is a multivariate Gaussian distribution:
\begin{align}
p(\vt{x}_{c} \mid \vt{y}_c)
& {}= \N{\vt{x}_{c} \mid \vt{A}_{c} \vt{y}_c,\tau_{c}^{-1} 
{\bf I}} \cr
& {}= \frac{\tau_{c}^{N_{c}/2}}{(2 \pi)^{N_{c}/2}} 
\exp \left( 
-\frac{\tau_{c}}{2} \norm{{\bf x}_{c} - {\bf A}_{c} {\bf y}_c}_2^2 \right), \label{eq:superres:cond}
\end{align}
where $\tau_{c}$ is the precision of the noise of LR image $c$ and
$\vt{A}_{c} \in \mathbb{R}^{N_{c} \times N}$ is the linear operator
mapping from HR to LR space. Dropping all terms that do not depend on
$\vt{y}_c$, the negative log-likelihood can be written as:
\begin{align}
-\ln p(\vt{x}_{c} \mid \vt{y}_c) = \frac{\tau_{c}}{2} \norm{{\bf x}_{c} - {\bf A}_{c} {\bf y}_c}_2^2 + \mathrm{const}.
\label{eq:superres:lncond}
\end{align}
The multivariate Gaussian distribution in \eqref{eq:superres:cond} is a
likelihood function that is already well-established in the
super-resolution literature
\citep{greenspan2001mri,shilling2009super,poot2010general}.

\begin{figure}[t]
\centering
\includegraphics[width=\columnwidth]{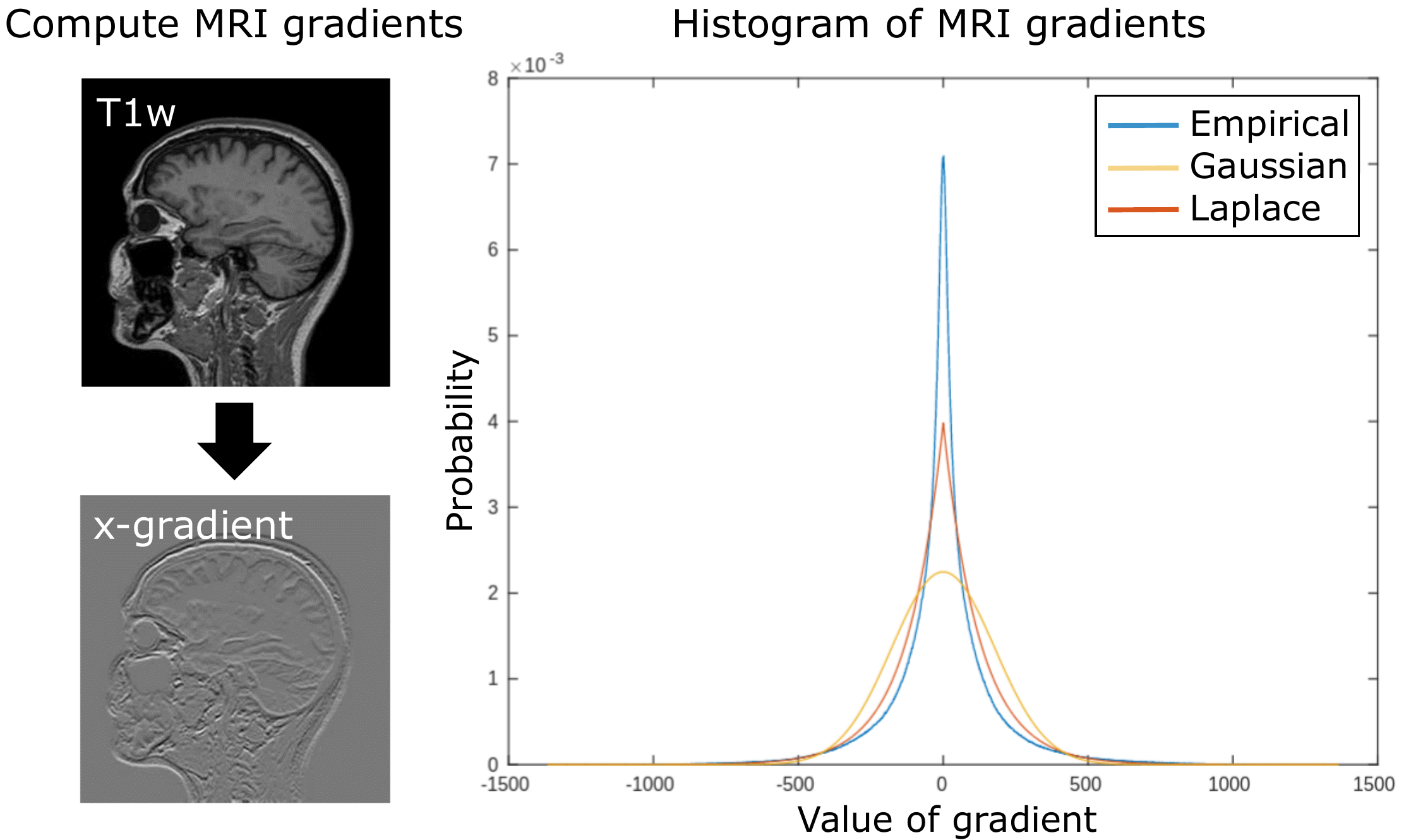}
\caption[Histogram of MR image gradients]{Empirical investigation of the
distribution of MR image gradients. By computing the $x$-, $y$- and
$z$-gradients of a MR image, and fitting a Gaussian (yellow) and a
Laplace distribution (red) to the histogram of these gradients (blue),
it can be seen that the Laplace distribution more accurately captures
the empirical distribution of MR image gradients.}
\label{fig:superres:Gauss-vs-Laplace}
\end{figure}

\subsubsection{Model Prior}

The prior probability should encode our belief about the unknown HR
images ($\vt{Y}$). Many types of priors have been devised for image
reconstruction problems. The most popular alternative is perhaps the
Tikhonov (or $\ell_2$ prior), that penalises the squared norm of some
image features:
\begin{align}
p(\vt{y}_c) = \N{\vt{y}_c\mid\vm{0},(\lambda_c \vt{L})^{-1}}, 
\end{align}
where $\lambda_c$ is a channel specific regularisation parameter and the
precision matrix ($\vt{L}$) is designed as to induce correlations
between image voxels. This type of prior probability favours images that
are smooth when the precision matrix encodes some differential operator
$\vt{L} = \vt{D}\tr\vt{D}$, so that:
\begin{align}
-\ln p(\vt{y}_c) = \frac{\lambda}{2} \norm{\vt{D}\vt{y}_c}_2^2 + 
\text{const}. 
\label{eq:superres:fot}
\end{align}
If the differential operator encodes a first-order derivative, then the
negative log-likelihood of this prior is known as a first-order Tikhonov
regulariser. The differential operator is in dimension $\vt{D} \inr{NG
\times N}$, where $G$ are the number of differential features.

The underlying assumption of the Tikhonov prior is that the gradients in
the HR image have a Gaussian distribution. However, by studying the
empirical gradient distribution of a high-resolution, noise-free MR
image we see that a Laplace distribution is more suitable (see Figure
\ref{fig:superres:Gauss-vs-Laplace}). Discarding terms that do not
depend on $\vt{y}_c$, this gives the following prior distribution:
\begin{align}
p(\vt{y}_c) &\propto \prod_{n=1}^{N} \exp \left(- \lambda_c
\norm{\vt{D}_n\vt{y}_c}_2\right), 
\label{eq:superres:laplace}\\
-\ln p(\vt{y}_c) &= \lambda_c \sum_{n=1}^N \norm{\vt{D}_n\vt{y}_c}_2 + \text{const}.
\label{eq:superres:tv}
\end{align}
where $\lambda_c$ is the inverse of the scale parameter of the Laplace
distribution, and the operator $\vt{D}_n \in \mathbb{R}^{G\times N}$
returns the gradients at the $n$th data point of $\vt{y}_c$. The log of
the Laplace distribution in \eqref{eq:superres:tv} is known as
(isotropic) total variation (TV) and is another popular method for
regularising image reconstruction problems \citep{rudin1992nonlinear}.
Rather than favouring smooth reconstructions, TV retains edges and
therefore leads to less blurry results. Here, to avoid biasing the
reconstruction, we extract both the forward and backward first-order
finite differences along each dimension, giving $G=6$ in 3D.

If we assume that the unknown HR images follow the distribution
in \eqref{eq:superres:laplace}, then we do not have any dependencies
between channels. Clearly this assumption is false for MR images of the
same subject, where most edges should be shared between channels. We
therefore propose using the multi-channel total variation (MTV)
functional \citep{sapiro1996anisotropic} as our prior probability:
\begin{align}
p(\vt{Y}) &\propto \prod_{n=1}^{N} \exp \left(- 
\sqrt{\sum_{c=1}^{C} \norm{\lambda_c \vt{D}_n\vt{y}_c}_2^2}\right), 
\label{eq:superres:mtvdist}\\
-\ln p(\vt{Y}) &= \sum_{n=1}^{N}  
\sqrt{\sum_{c=1}^{C} \norm{\lambda_c \vt{D}_n\vt{y}_c}_2^2} + 
\text{const}.
\end{align}
The summation over channels ($C$) inside of the square root ensures that
the channels are `mixed', making the assumption that the MR images have
large smooth regions and a few sharp edges, in similar places. Note that
TV is a special case of MTV, when $C=1$.

\begin{figure*}[t]
\centering
\includegraphics[width=0.8\textwidth]{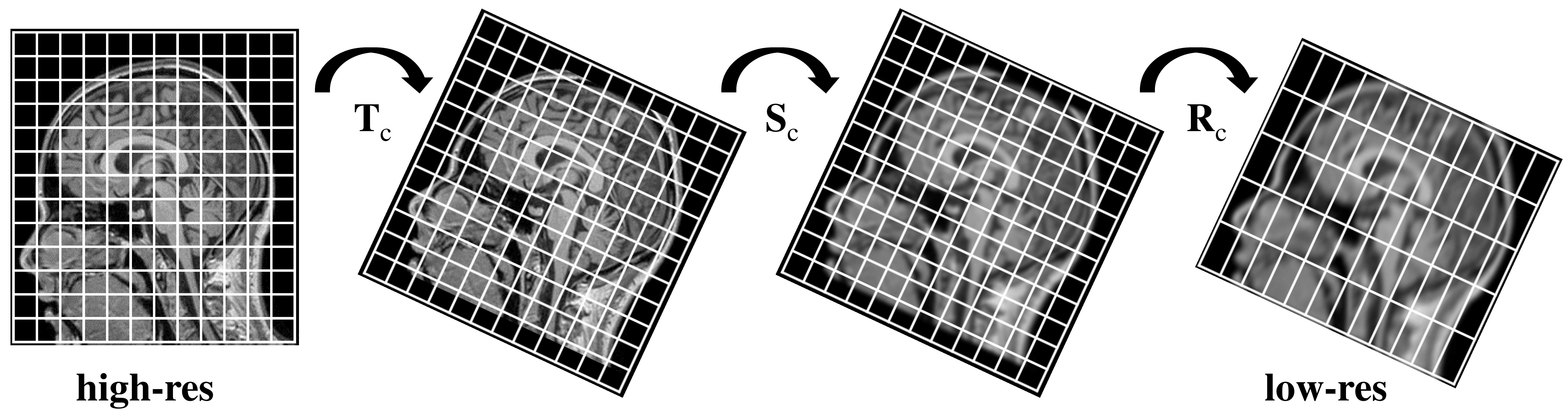}
\caption[Illustration of super-resolution forward model]{Illustration of
the super-resolution forward model ($\vt{A}_c$). The operator $\vt{T}_c$
resamples from the HR image's FOV to the LR image's FOV, keeping the
voxel size of the HR image. The operator $\vt{S}_c$ simulates the slice
profile of the MRI acquisition. The operator $\vt{R}_c$ performs the
down-sampling operation from HR to LR space.}
\label{fig:superres:projmat2}
\end{figure*}

\subsubsection{Model Parameters}
\label{sec:methods:model:parameters}

Defining the generative model has given us a set of parameters: the
projection matrices ($\vt{A}$), noise precisions ($\tau$) and prior
parameters ($\lambda$). These parameters are all image-specific and it
is critical that we set these parameters in a principled way for our
model to generalise -- it is not feasible to expect users to do manual
tuning. This section gives more detail about how the model parameters
were chosen.

\begin{figure*}[t]
\centering
\subfloat[]{\includegraphics[width=0.49\textwidth]
{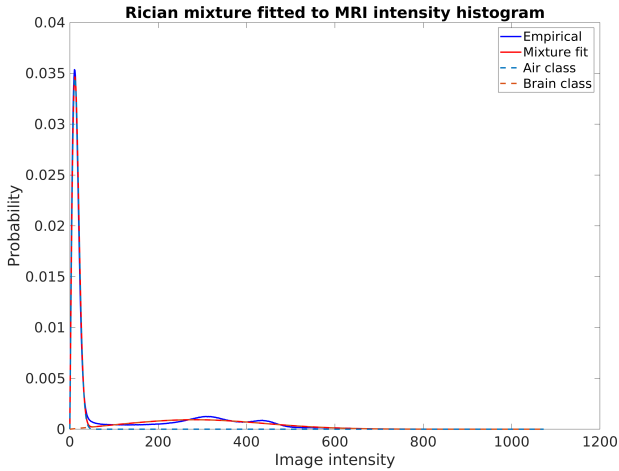}\label{fig:superres:tau}} \hfil
\subfloat[]{\raisebox{0pt}{\includegraphics[width=0.49\textwidth]
{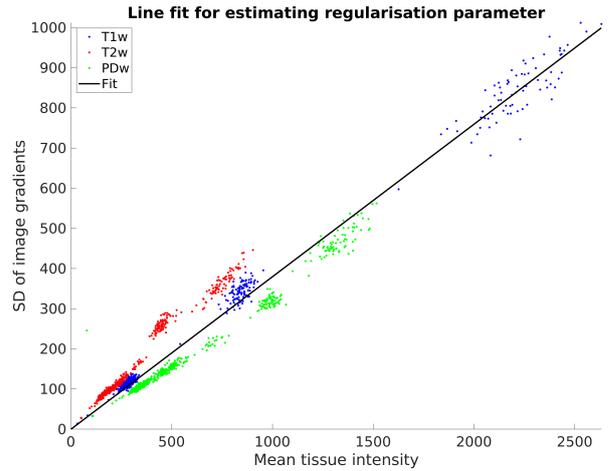}\label{fig:superres:lambda}}}
\caption[Estimating model parameters for the super-resoution
model]{Estimating model parameters for the super-resolution model. (a) A
two-class Rician mixture model was fit to the intensity histogram of
each MR image. The image noise was computed as the variance of the air
class. Note that image intensities in MRI are non-quantitative. (b) A
straight line was fit between the mean tissue intensities and the
standard deviations of image gradients, computed from the IXI dataset;
its coefficient gives us the value of the parameter $k_{\lambda}$.}
\end{figure*}

\begin{description}
\item[Projection matrices:] 

The projection matrix ($\vt{A}_c$) is a linear mapping from HR to LR
image space. It should reproduce the slice-selection process of the MRI
scanner. In this work we design the projection matrix as three linear
operators, applied in succession as: \begin{align} \vt{A}_c = \vt{R}_c
\vt{S}_c \vt{T}_c. \end{align} The operator $\vt{T}_c$ resamples from
the HR image's field of view (FOV) to the LR image's FOV, but keeping
the voxel size of the HR image. If the slice orientation is at an angle
with respect to the HR grid, the corresponding rotation is accounted
for. Furthermore, to improve numerical properties of the projection
operator, we slightly increase the FOV of the resampled image, which is
then adjusted for in the $\vt{R}_c$ operator.

The operator $\vt{S}_c$ should simulate the slice profile of the MRI
acquisition. The slice profile depends on the shape of the
radio-frequency (RF) pulse applied during slice selection. This RF pulse
can vary a lot from one sequence to the next. Therefore, there is not a
single slice profile that suits all acquisitions \citep{liu2002actual}.
Here, we make the assumption that the slice profile is
Gaussian\footnote{Our implementation of the projection matrix however,
gives a user the option to easily change the slice profile assumption by
simply changing the convolution kernel.}. Applying the $\vt{T}_c$
operator before the Gaussian convolution ensures that the kernel is
applied in the correct directions. We set the full width at half maximum
(FWHM) of the Gaussian kernel to zero in the in-plane directions and to
the width of the thick-slice in the thick-slice direction. We
additionally modulate the thick-slice direction FWHM by subtracting a
slice gap. We computed an estimate of this slice-gap from 29,026
patients MRIs\footnote{From the DICOM representation of a MR image,
whose header contain both the field \texttt{Spacing Between Slices}
(0018, 0088) and \texttt{Slice Thickness} (0018, 0050), the slice gap
can be computed as the \texttt{Spacing Between Slices} minus the
\texttt{Slice Thickness}.}. The estimate we obtain for the slice gap is
one third of the width of the slice thickness. This information could be
read from the DICOM header of the images. But as our tool works on NIfTI
data, this information may not be present, as it can be lost during
DICOM-to-NIfTI conversion. The slice gap can however be provided, when
available.

The operator $\vt{R}_c$ performs the down-sampling operation from HR to
LR space. The process of applying $\vt{A}_c$ to a HR image is
illustrated in Figure \ref{fig:superres:projmat2}.

\item[Noise precision:] 

MR scans are usually reconstructed as the magnitude of an image that was
originally complex. The Gaussian noise model in
\eqref{eq:superres:noisemodel} is therefore just an approximation to a
Rician noise model. Hence, we want to estimate the amount of Rician
noise in each observed LR image ($\tau_c$). We do so by fitting a
mixture of two Rician distributions to the intensity histogram of each
MR scan \citep{ashburner2013symmetric}. We then calculate the precision
of the image noise ($\tau_c$) from the class with the smallest
noncentrality parameter, which should correspond to air. An example fit
is shown in Figure \ref{fig:superres:tau}.

\item[Regularisation parameters:] 

Each unknown HR image has a corresponding Regularisation parameter
($\lambda_c$). First, note that that gradient values are, in general,
correlated with intensity values; the regularisation parameter should
therefore be set with respect to the mean intensity in a channel. We
observed empirically a linear relationship between the standard
deviation of gradient magnitudes $\sigma_c$ and the mean tissue
intensity of an image $\mu_c$. We computed the mean tissue intensity of
1,728 scans from the IXI dataset by fitting a two-class Rician
distribution and taking the noncentrality parameter of its non-air
component, and regressed the standard deviation of the first-order
gradients against it. We obtained the value $k_{\lambda} = \sigma/\mu =
4.67$; the fit is shown in Figure \ref{fig:superres:lambda}. Since the
variance of a Laplace distribution relates to the scale parameter
through the relation $\sigma^2 = 2/\lambda^2$, we can set the parameter
$\lambda_c$ according to:
\begin{align}
\lambda_c = \frac{\sqrt{2}}{\sigma_c} = \frac{\sqrt{2}}{k_{\lambda} 
\mu_c}.
\end{align}

\end{description}


\subsection{Model Optimisation}

With all the individual components of the model defined, the negative 
log posterior probability can be written as:
\begin{align}
- \ln \left(p(\bm{\mathcal{X}}\mid{\bf Y})~p({\bf Y})\right)
=& \sum_{c=1}^C \frac{\tau_{c}}{2} 
\norm{{\bf x}_{c} - {\bf A}_{c} {\bf y}_c}_2^2 \cr
&+ \sum_{n=1}^{N}  \sqrt{\sum_{c=1}^C \norm{\lambda_c {\bf D}_n 
{\bf y}_{c}}_2^2} \cr
&+ \text{const},
\label{eq:superres:objfun}
\end{align}
where the first term on the right-hand side is known as the data term
and the second term as the penalty term. The expression in
\eqref{eq:superres:objfun} is what we want to minimise to obtain the $C$
super-resolved images:
\begin{align}
\argmin_{\bf Y} \left\{ - \ln  
p(\bm{\mathcal{X}}\mid{\bf Y})~p({\bf Y}) \right\}.
\label{eq:superres:optprob}
\end{align}
This optimisation problem is hard to solve because the MTV penalty term
is non-differentiable (nonsmooth). It is, however, convex with a global
optimum.

A change of variables allows the unconstrained minimisation problem in
\eqref{eq:superres:optprob} to be rewritten as a constrained
minimisation:
\begin{align}
\begin{split}
\min_{\vt{Y}}   \qquad &\sum_{c=1}^C \frac{\tau_{c}}{2} 
\norm{{\bf x}_{c} - {\bf A}_{c} {\bf y}_c}_2^2 + \sum_{n=1}^{N}  
\sqrt{\sum_{c=1}^C \norm{\vt{z}_{nc}}_2^2} \\
\text{s.t.} \qquad &\lambda_c {\bf D}_n 
{\bf y}_{c} = \vt{z}_{nc}, \quad \text{for all $n$ and 
$c$},
\label{eq:superres:objfunconst}
\end{split}
\end{align}
where the smooth data term and the nonsmooth penalty term have been
decoupled. The constrained form of \eqref{eq:superres:optprob} can now
be solved efficiently by an algorithm known as alternating direction
method of multipliers (ADMM).

\subsubsection{Alternating Direction Method of Multipliers}

ADMM is part of a class of optimisation methods called proximal
algorithms \citep{boyd2011distributed}. In short, an augmented
Lagrangian is formulated from a constrained minimisation problem. This
Lagrangian is then minimised in an alternating fashion until a
convergence criterion is met. Many methods exist for solving TV
problems. We chose an ADMM algorithm because it is straightforward
to implement and suitable for the type of optimisation problem in
\eqref{eq:superres:optprob}.

Deriving the general form of the ADMM algorithm starts with an
unconstrained minimisation problem:
\begin{align}
\begin{split}
\min_{\vt{y}}   \qquad &\text{data}(\vt{y}) + \text{penalty}(\vt{y}),
\end{split}
\end{align}
which is equivalent to the constrained problem:
\begin{align}
\begin{split}
\min_{\vt{y},\vt{z}}   \qquad &\text{data}(\vt{y}) + 
\text{penalty}(\vt{z}) 
\cr
\text{s.t.} \qquad &\vt{F}_d\vt{y} + \vt{F}_p\vt{z} = \vt{d},
\label{eq:superres:admmobjfun}
\end{split}
\end{align}
with constraints defined by $\vt{F}_d$, $\vt{F}_p$ and $\vt{d}$. Note
how the optimisation problem in \eqref{eq:superres:admmobjfun} has the
same form as the super-resolution minimisation in
\eqref{eq:superres:objfunconst}.

From \eqref{eq:superres:admmobjfun}, the augmented Lagrangian can be 
formulated as:
\begin{align}
\mathcal{L}_{\rho} (\vt{y},\vt{z},\vt{w}) ={}& \text{data}(\vt{y}) + 
\text{penalty}(\vt{z}) \cr
&+ \vt{w}\tr (\vt{F}_d\vt{y} + \vt{F}_p\vt{z} - 
\vt{d}) 
\cr
&+ \frac{\rho}{2} \norm{\vt{F}_d\vt{y} + \vt{F}_p\vt{z} - \vt{d}}_2^2,
\label{eq:superres:auglag}
\end{align}
where $\vt{w} \in \mathbb{R}^{P}$ holds the Lagrange multipliers and
$\rho > 0$ is a descent step-size. From the augmented Lagrangian in
\eqref{eq:superres:auglag}, the ADMM updates are given as:
\begin{align}
\vt{y}^{k + 1} &\coloneqq \argmin_{\vt{y}} \mathcal{L}_{\rho} 
(\vt{y},\vt{z}^k,\vt{w}^k), 
\label{eq:superres:admmupdatey}\\
\vt{z}^{k + 1} &\coloneqq \argmin_{\vt{z}} \mathcal{L}_{\rho} 
(\vt{y}^{k+1},\vt{z},\vt{w}^k), 
\label{eq:superres:admmupdatez}\\
\vt{w}^{k + 1} &\coloneqq \vt{w}^k + \rho\cdot(\vt{F}_d\vt{y}^{k + 1} 
+ 
\vt{F}_p \vt{z}^{k + 1} - \vt{d}),
\label{eq:superres:admmupdatew}
\end{align}
where \eqref{eq:superres:admmupdatey} and
\eqref{eq:superres:admmupdatez} are known as the proximal operators at
parameter $\rho$ for the data and penalty term, respectively. These ADMM
updates are usually iterated until some convergence criteria is
fulfilled.

\subsubsection{ADMM updates for the super-resolution model}

In this section we derive the ADMM updates for our super-resolution
model. We will work with two tensors $\vm{\mathcal{Z}}$ and
$\vm{\mathcal{W}}$, which are both of dimensions $N \times C \times G$.
We extract column vectors from these tensors, where the subscript of a
vector indicates its length: $\vt{z}_n =
\text{vec}(\vm{\mathcal{Z}}[n,:,:])$, $\vt{z}_c =
\text{vec}(\vm{\mathcal{Z}}[:,c,:])$, $\vt{z}_{nc} =
\text{vec}(\vm{\mathcal{Z}}[n,c,:])$, $\vt{w}_n =
\text{vec}(\vm{\mathcal{W}}[n,:,:])$ and $\vt{w}_c =
\text{vec}(\vm{\mathcal{W}}[:,c,:])$. With the notations introduced we
formulate the augmented Lagrangian in \eqref{eq:superres:auglag} from
\eqref{eq:superres:objfunconst} as:
\begin{align}
\mathcal{L}_{\rho} (\vt{Y},\vm{\mathcal{Z}},\vm{\mathcal{W}}) =& 
\sum_{c=1}^C \frac{\tau_{c}}{2} \norm{{\bf x}_{c} - {\bf A}_{c} {\bf
y}_c}_2^2 + \sum_{n=1}^{N}  \norm{\vt{z}_{n}}_2
\cr
&+ \sum_{c=1}^C \vt{w}_c\tr
(\lambda_c \vt{D}\vt{y}_c -
\vt{z}_c) \cr
&+ \frac{\rho}{2} \sum_{c=1}^C \norm{\lambda_c \vt{D}\vt{y}_c - 
\vt{z}_c}_2^2,
\label{eq:superres:auglagsr}
\end{align}
where we have made use of the fact that, for a fixed $n$:
\begin{align}
\sqrt{\sum_{c} \norm{\vt{z}_{nc}}_2^2} = \sqrt{\sum_c \sum_g 
(z_{ncg})^2} = \norm{\vt{z}_{n}}_2.
\end{align}
From the augmented Lagrangian in \eqref{eq:superres:auglagsr} the ADMM 
updates can be derived as:
\begin{align}
\vt{y}_c^{k + 1} \coloneqq \argmin_{\vt{y}_c} \Big(&
\frac{\tau_{c}}{2} \norm{{\bf x}_{c} - {\bf A}_{c} {\bf y}_c}_2^2 +
\lambda_c \vt{w}_c^k\tr \vt{D}\vt{y}_c  \cr
&+ \frac{\rho}{2} \norm{\lambda_c \vt{D}\vt{y}_c - 
\vt{z}_c^k}_2^2\Big), \label{eq:superres:yupdsr}\\
\vt{z}_n^{k + 1} \coloneqq \argmin_{\vt{z}_n} \Big(&
\norm{\vt{z}_{n}}_2 -  
\vt{w}_n^k\tr
\vt{z}_n \cr
&+ 
\frac{\rho}{2}\norm{\lambda_c \vt{D}_n\vt{y}_c^{k + 1} -
\vt{z}_n}_2^2\Big).
\label{eq:superres:zupdsr}
\end{align}
The updates in \eqref{eq:superres:yupdsr} and \eqref{eq:superres:zupdsr}
give $C$ optimisation problems for the HR images $\vt{Y}$ and $N$
optimisation problems for the variables $\vm{\mathcal{Z}}$. From
\eqref{eq:superres:admmupdatew} the estimate of the Lagrange multiplier
is:
\begin{align}
\vt{w}_c^{k + 1} &\coloneqq \vt{w}_c^k + 
\rho\cdot(\lambda_c 
\vt{D}\vt{y}_c^{k+1} - 
\vt{z}_c^{k +1}), \quad \text{for all $c$}.
\label{eq:superres:admmupdateww}
\end{align}

\paragraph{ADMM update for $\vt{Y}$:} The optimisation problem in
\eqref{eq:superres:yupdsr} is simply regularised least-squares with a
closed-form solution given by:
\begin{align}
{\bf y}_{c}^{k+1} =& \bigg( \tau_{c} \vt{A}_{c}\tr {\bf
A}_{c}  + \rho \lambda_c^2{\bf D}\tr {\bf D} \bigg)^{-1} \cr
&
\bigg(\tau_{c} {\bf A}_{c}\tr {\bf x}_{c} - \lambda_c {\bf D}\tr (  
{\bf w}_{c} - \rho {\bf
z}_{c}) \bigg).
\label{eq:superres:linsys}
\end{align}
This system is too large to be inverted directly, and a conjugate
gradient method is often used as an alternative
\citep{hestenes1952methods}. Here, for increased speed and convergence,
we instead use a Newton's method. Writing the objective function as:
\begin{align}
\mathcal{L}(\vt{y}_c) =& \frac{\tau_{c}}{2} \norm{{\bf x}_{c} - {\bf 
A}_{c} {\bf y}_c}_2^2 +
\lambda_c \vt{w}_c\tr
\vt{D}\vt{y}_c \cr
&+ \frac{\rho}{2} \norm{\lambda_c 
\vt{D}\vt{y}_c - 
\vt{z}_c}_2^2,
\end{align}
we get its gradient as:
\begin{align}
\frac{\partial \mathcal{L}(\vt{y}_c)}{\partial \vt{y}_c} =& (\tau_c
\vt{A}_c\tr\vt{A}_c + \rho \lambda_c^2 \vt{D}\tr\vt{D})\vt{y}_c \cr
&+
\lambda_c \vt{D}\tr (\vt{w}_c - \rho \vt{z}_c) - \tau_c \vt{A}_c\tr
\vt{x}_c,
\end{align}
and Hessian as:
\begin{align}
\frac{\partial^2 \mathcal{L}(\vt{y}_c)}{\partial \vt{y}_c \partial 
\vt{y}_c\tr}
= (\tau_c \vt{A}_c\tr\vt{A}_c + \rho \lambda_c^2 \vt{D}\tr\vt{D}).
\end{align}
Since the problem is quadratic, Newton's method would solve it in one
step, which is equivalent to computing the closed-form solution in
\eqref{eq:superres:linsys}. However, obtaining the full Hessian is
computationally difficult. We therefore replace the true Hessian by a
majorising matrix:
\begin{align}
\vt{H}_c = \text{diag}(\tau_c \vt{A}_c\tr\vt{A}_c\mathds{1})  + \rho
\lambda_c^2 \vt{D}\tr\vt{D},
\end{align}
where $\mathds{1} \inr{N}$ is a vector of ones. This approximation,
which can be inverted in linear time using a multigrid method
\citep{ashburner2007fast}, is more positive-definite than the true
Hessian in the Loewner ordering sense (see Lemma S.3 in
\cite{chun2017convolutional}) and therefore ensures convergence. This
gives us the following update step:
\begin{align}
\vt{y}_c^{k+1} = \vt{y}_c - \vt{H}_c^{-1} \frac{\partial 
\mathcal{L}(\vt{y}_c)}{\partial \vt{y}_c}.
\label{eq:superres:gny}
\end{align}

\paragraph{ADMM update for $\vm{\mathcal{Z}}$:} By some algebraic
manipulations we can rewrite \eqref{eq:superres:zupdsr} in the
equivalent form:
\begin{align}
\vt{z}_n^{k+1} = \argmin_{\vt{z}_n} \Big(&\frac{\rho}{2} 
\norm{\vt{z}_n - 
\left(\frac{1}{\rho}\vt{w}_n + \lambda_c 
\vt{D}_n\vt{y}_c\right)}_2^2 \cr
& +
\norm{\vt{z}_{n}}_2 + \text{const}\Big),
\label{eq:superres:superresz}
\end{align}
where the constant contains terms that do not depend on $\vt{z}_n$. This
gives us an optimisation problem with an $\ell_2$ data term and an
$\ell_1$ penalty term. A more general formulation of this problem is:
\begin{align}
\argmin_{\vt{s}} \Big( \frac{\beta}{2} \norm{\vt{s} 
- \vt{t}}_2^2 + \alpha \norm{\vt{s}}_2\Big),
\label{eq:superres:softth1}
\end{align}
which has a closed-form solution given by (proof given in, \emph{e.g.}, 
\cite{yang2009fast}):
\begin{align}
\vt{s}(\vt{t}) = \text{max}\Bigg\{ \norm{\vt{t}}_2 - 
\frac{\alpha}{\beta} , 0\Bigg\} \odot \frac{\vt{t}}{\norm{\vt{t}}_2},
\label{eq:superres:softth2}
\end{align}
where $\odot$ denotes the Hadamard product. The solution to the
optimisation problem in \eqref{eq:superres:superresz} is therefore given
by:
\begin{align}
\vt{z}_n^{k+1} ={} & \text{max}\Bigg\{ \norm{\frac{1}{\rho}\vt{w}_n + 
\lambda_c 
\vt{D}_n\vt{y}_c}_2 - 
\frac{1}{\rho} ,0\Bigg\} \cr
&\odot \frac{\frac{1}{\rho}\vt{w}_n + 
\lambda_c 
\vt{D}_n\vt{y}_c}{\norm{\frac{1}{\rho}\vt{w}_n + \lambda_c 
\vt{D}_n\vt{y}_c}_2}.
\label{eq:superres:solz}
\end{align}

\begin{algorithm}[t]
\caption{Multimodal super-resolution}
\begin{algorithmic}[1]
\State Coregister input images $(\vt{X})$.
\State Estimate model parameters $(\vm{\tau},\vm{\lambda})$.
\State Initialise variables to zero 
$(\vt{Y},\vm{\mathcal{Z}},\vm{\mathcal{W}})$.    
\While{not coverged}
\For{$c=\{1,\ldots,C\}$}
\State \textit{\# Loop over channels (distributed)}
\State Compute $\vt{y}_c^{k + 1}$ by \eqref{eq:superres:gny}, when 
given $\vt{z}_c^{k}$ and $\vt{w}_c^{k}$.
\EndFor
\For{$n=\{1,\ldots,N\}$}
\State \textit{\# Loop over HR voxels}
\State Compute $\vt{z}_n^{k + 1}$ by \eqref{eq:superres:solz}, when 
given 
$\vt{y}_n^{k + 1}$ and $\vt{w}_n^{k}$.
\EndFor
\For{$c=\{1,\ldots,C\}$}
\State \textit{\# Loop over channels}
\State Compute $\vt{w}_c^{k + 1}$ by \eqref{eq:superres:admmupdatew}, 
when given 
$\vt{y}_c^{k + 1}$ and $\vt{z}_c^{k + 1}$.
\EndFor
\EndWhile
\end{algorithmic}
\label{alg:superres:1}
\end{algorithm}

\subsection{Implementation Details}

Algorithm \ref{alg:superres:1} shows the steps for super-resolving a set
of thick-sliced patient scans with the proposed model. The process is
computationally intensive and we take care to provide an efficient
implementation. The software is written in a mixture of MATLAB and C
code. There are large memory requirements, which are likely to exceed
the random-access memory (RAM) of some workstations. To save memory, all
the computations are therefore performed using single precision floating
point. This lower precision has negligible effect on the numerical
stability. We solve for $\vt{Y}$ efficiently using a multigrid technique
\citep{ashburner2007fast}. The Newton update can be iterated over to get
a closer fit to the solution. In this work, however, a single update is
used to reduce computational time. Furthermore, the loop solving for
each $\vt{Y}$ is distributed, so that a separate process handles each
iteration. Finding the optimal step-size $\rho$ is an open-problem
\citep{dohmatob2014benchmarking}, and though under mild conditions ADMM
converges for any value of $\rho$, the convergence rate depends on
$\rho$. Here, we use the heuristic:
\begin{align}
\rho = 
\frac{\sqrt{\text{mean}(\{\lambda_c\}_{c=1}^C)}}{\text{mean}(\{\tau_c\}_{c=1}^C)},
\end{align}
which we observe empirically gives good convergence properties.
Furthermore, algorithm convergence is defined by the relative change in
objective value: $2\cdot(l^{k} - l^{k + 1})/(l^{k} + l^{k + 1})$, being 
less
than $10^{-4}$. The objective value ($l$) is computed from
\eqref{eq:superres:objfun}.
Finally, we define the HR images' FOV as the bounding box that contains
all LR images' FOV, and we set the voxel size of the HR images to 1 mm
isotropic.

\section{Evaluation}

The aim of this section is to assess the effectiveness of the MTV prior,
compared with classically used regularisers for super-resolving
multi-contrast MR datasets, and to illustrate the proposed model's
ability to process a large clinical dataset.

\begin{figure*}
\centering
\includegraphics[width=0.8\textwidth]{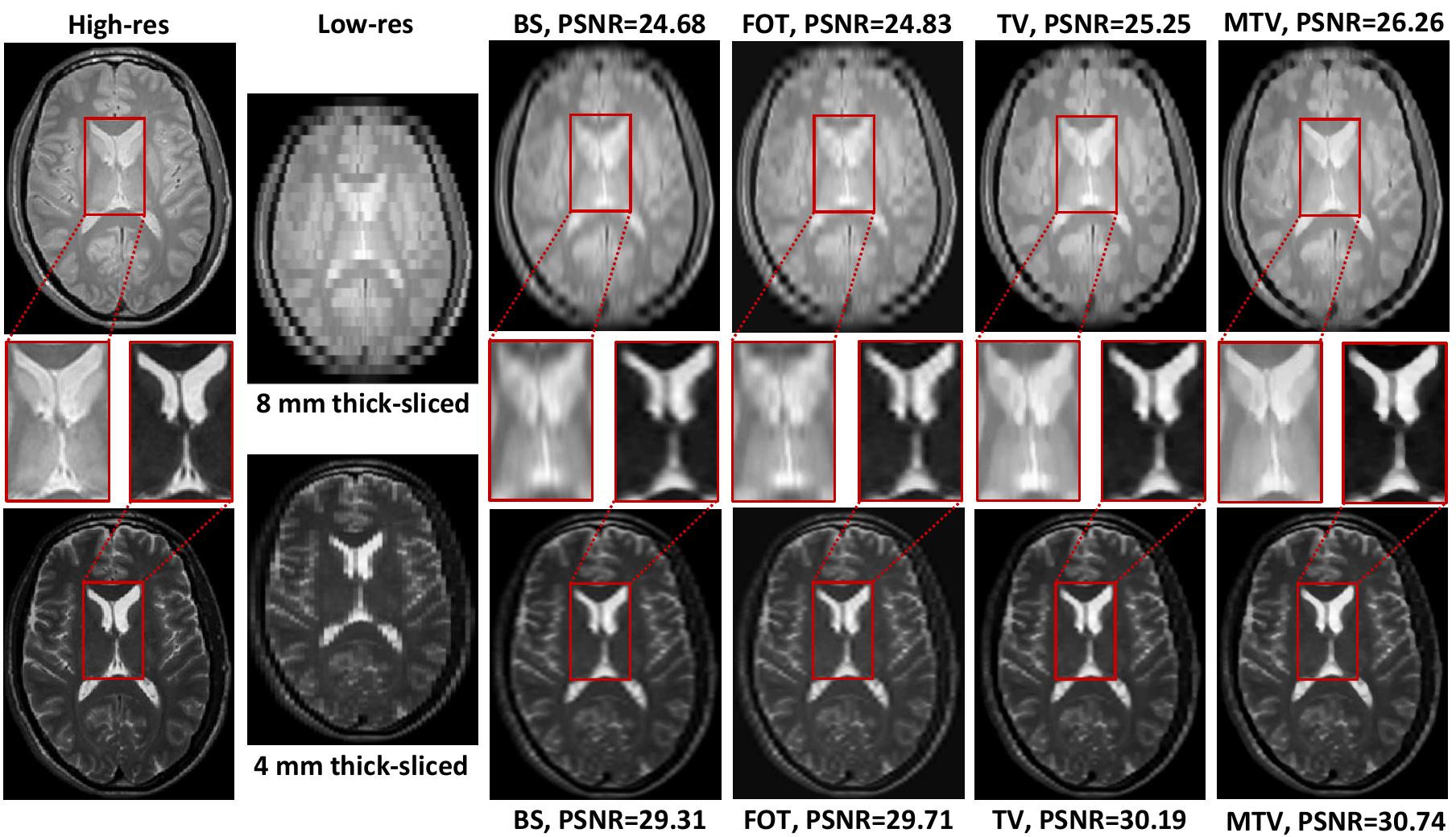}
\caption[Example of simulating and super-resolving from the IXI
dataset]{Example of simulating and super-resolving from the IXI dataset.
Two LR images were simulated from two HR images (PDw and
T2w). The HR images were then reconstructed using four methods: 
BS, FOT and TV are single-channel SR techniques, capable of combining 
LR images of only one channel; MTV on the other hand, a 
multi-channel method, combines information from both channels.}
\label{fig:superres:recs}
\end{figure*}

\subsection{High-resolution IXI dataset}

The IXI
dataset\footnote{\url{http://brain-development.org/ixi-dataset/}}
contains multimodal MR volumes, with T1w, T2w and PDw images, of 576
healthy subjects, acquired on 1.5T and 3T systems at three different
centres. All images have close to 1 mm isotropic resolution. This
dataset was used to (1) validate the robustness of the noise variance
estimation by adding known amounts of Rician noise and estimating it;
(2) compare our heuristic for setting regularisation parameters, with
optimal values obtained by grid-search; (3) compare the efficiency of
two iterative methods (conjugate gradient and approximate Newton) for
solving a quadratic optimisation problem; (4) compare four different
methods for super-resolving MRIs: 4th order b-spline (BS) interpolation,
first-order Tikhonov (FOT), TV and MTV.

LR images were generated by applying the forward projection operator
($\vt{A}$) to HR images, such that LR images had thick slices in one
direction. Thick-slice directions were picked randomly, but were always
orthogonal across contrasts (\emph{e.g.}, axial for T1w, coronal for
T2w, sagittal for PDw). Figure \ref{fig:superres:recs} shows an example
of simulated LR images from HR images. Comparisons between methods are
based on the root-mean-square error (RMSE):
\begin{align}
\operatorname{RMSE} = \sqrt{\sum_{n=1}^N 
\left(y_{n}^{\text{recon}} - y_{n}^{\text{ref}}\right)^2},
\end{align}
and the peak signal-to-noise ratio (PSNR):
\begin{align}
\operatorname{PSNR} = 
20 
\log_{10}\left(\frac{\max\left(\vt{y}^{\text{ref}}\right)}{\operatorname{RMSE}}\right),
\end{align}
two widely used metrics for image reconstruction. These metrics were
computed for each contrast. We also computed Dice scores between GM and
WM segmentations obtained by applying SPM12's unified segmentation (with
default parameters) to the super-resolved and HR images. Segmentation is
a typical processing task in neuroimaging as it allows for morphometric
analyses to be conducted.

FOT, defined in \eqref{eq:superres:fot} and TV, defined in
\eqref{eq:superres:tv}, were implemented within the same framework as
MTV. Therefore, they used the same projection matrices and parameters.
B-spline interpolation is not technically a super-resolution technique,
but is often used in practice to reslice low-resolution images prior to
automated processing. For b-spline interpolation, if multiple LR images
of the same contrast are available, they are simply averaged.

\begin{description}

\item[Noise Precision:]

All 1,728 IXI scans were used in this experiment. A known amount of
Rician noise, as a percentage of the mean image intensity (1\%, 2.5\%,
5\%, 10\%), was added to each image. The two-class Rician mixture model
was then fit to the intensity histogram of the noisy images. The
variance of the cluster with the lowest noncentrality parameter was used
as an estimate of the image noise percentage. Table
\ref{tab:superres:noise_estimate} shows the mean and standard deviation,
across subjects, of the estimated noise variance percentage. These
results show that the Rician mixture model can accurately estimate a
wide range of noise levels. Note that there is already Rician noise
present in the images. However, the amount of Rician noise that is added
is an order of magnitude greater than the intrinsic noise.

\begin{table*}[t]
\fontsize{8}{7.2}\selectfont
\centering
\caption[Validating the estimate of the noise precision
parameter]{Validating the estimate of the noise precision parameter (for
1,728 subjects). Showing the simulated noise percentages, and the noise
percentages estimated by fitting the two class Rician mixture model.
Shown as mean$\pm$std.}
\begin{widetable}{\textwidth}{*{5}{c}} \toprule
Ground-truth (\%)&1&2.5&5&10\\\midrule
Estimated (\%)& $1.10\pm0.76$ & $2.54\pm0.51$ 
&$5.23\pm1.01$ &$10.31\pm2.49$ \\\bottomrule
\end{widetable}
\label{tab:superres:noise_estimate}
\end{table*}

\item[Regularisation Parameter:]

Four images from different subjects and contrasts (T1w, T2w, PDw and
diffusion-weighted (DW)), were used to validate whether the heuristic
devised in section \ref{sec:methods:model:parameters} to estimate the
prior parameter ($\lambda$) yields proper regularisation values. LR
images with seven millimetre slice-thickness were generated and a
grid-search over the regularisation parameter was performed, (in the
range $\left[10^{-4},1\right]$, with step size $0.2$). For each value,
PSNR between the resulting MTV super-resolved images and the known
ground-truth was computed. A DW image was included to investigate
whether the heuristic generalises to contrasts not part of the training
set. The result of each grid-search, with the corresponding heuristic
estimates marked by crosses, can be seen in Figure
\ref{fig:superres:est-lam}. This experiment shows that the method for
estimating the regularisation parameter allows near-optimal
reconstructions to be produced, even for unseen contrasts.

\begin{figure}[t]
\centering  
\includegraphics[width=\columnwidth]{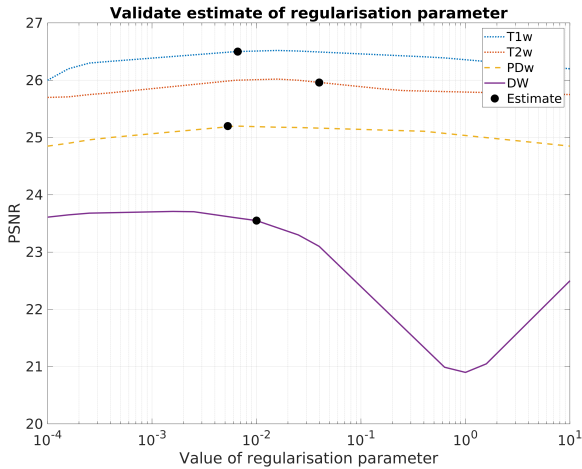}
\caption[Validation of estimating the regularisation
parameter]{Validation of estimating the regularisation parameter. A
grid-search over the regularisation parameter ($\lambda$) was performed
for four different contrasts, and PSNR between the reconstruction and
reference image was computed. The regularisation parameter obtained
heuristically is marked by a dot.}
\label{fig:superres:est-lam}
\end{figure}

\begin{figure}[t]
\centering  
\includegraphics[width=\columnwidth]{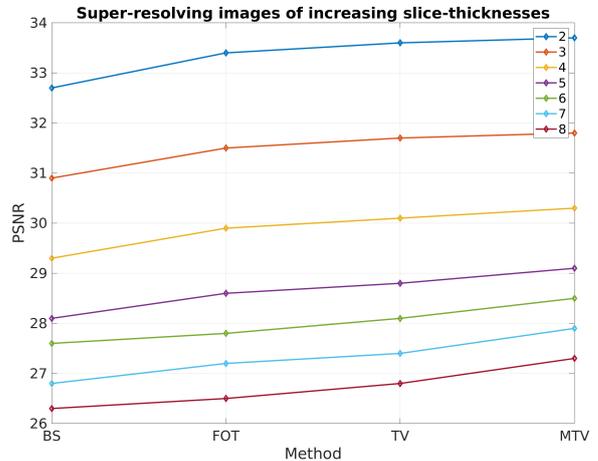}
\caption[Super-resolving images of different
slice-thicknesses]{Super-resolving images of different slice-thicknesses
(for 20 subjects). As the simulated slice-thickness increases (top to
bottom), MTV super-resolution results improve favourably compared with
the single-channel methods (BS, FOT, TV).}
\label{fig:superres:mtvvsoth}
\end{figure}

    
\item[Approximate Newton Solver:]

The update-step for $\vt{Y}$ entails solving a large linear system,
which is often done iteratively using the conjugate gradient method
\citep{hestenes1952methods}. Here, we show that, for this particular
problem, an approximate Newton method (based on a majoriser of the full
Hessian and solved with a multigrid algorithm) converges faster than the
conjugate gradient method. T1w, T2w and PDw image were simulated, with
6 mm slice thicknesses and different thick-slice directions. The MTV
super-resolution algorithm was then run, where the update for $\vt{Y}$
was solved either using the conjugate gradient or the approximate Newton
method. Both solution method were iterated for a fixed number of
iterations (50). Figure \ref{fig:superres:cg_vs_gn} shows the evolution
of the negative log-likelihood over computation time. This total
computation time takes into account both the number of iterations and
the computation time per iteration. The approximate Newton solver
converges faster than the conjugate gradient solver. Note that we did
not use pre-conditioning, which may have led to increased conjugate
gradient solver performance.

\begin{figure}[t]
\centering 
\includegraphics[width=\columnwidth]{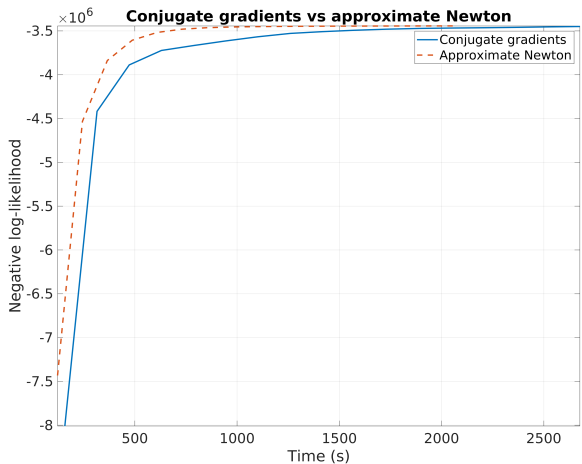}
\caption[Conjugate gradients vs approximate Newton for solving the
super-resolution problem]{Conjugate gradients vs approximate Newton for
solving the super-resolution problem. Time in seconds is plotted
against negative log-likelihood. It can be seen that the approximate
Newton method has faster convergence.}
\label{fig:superres:cg_vs_gn}
\end{figure}



\item[Multi-channel Super-Resolution:]

All IXI subjects were used to simulate LR images. For each subject, the
slice direction and thickness (between two and eight millimetre) were
chosen at random. The simulated LR images were super-resolved using BS,
FOT, TV and MTV. Among these, only MTV makes joint use of information
distributed across contrasts. Example reconstructions obtained with each
method can be seen in Figure \ref{fig:superres:recs}.

For each contrast, PSNR was computed between the reference and
super-resolved images. Table \ref{tab:superres:psnr} shows the average
PSNR and MTV obtained the greatest mean and lowest standard deviation.
Figure \ref{fig:superres:mtvvsoth} additionally shows average PSNRs for
different slice thicknesses, in which MTV once again performs
favourably.

Reference and super-resolved images were also segmented into GM and WM
and cerebrospinal fluid (CSF) using SPM12, with the reference
segmentation considered as ground-truth, and the Dice coefficient was
computed. The results can be seen in Table \ref{tab:superres:ds} where
MTV obtained the highest Dice score.

\begin{table*}[t]
\fontsize{8}{7.2}\selectfont
\centering
\caption[Results for single- vs multi-channel super-resolution]{Results
comparing single- vs multi-channel super-resolution. BS, FOT and TV are
all single-channel techniques, MTV is multi-channel. PSNRs are shown as
mean$\pm$std.}
\begin{widetable}{\textwidth}{*{5}{c}} \toprule
Channel & BS & FOT & TV & MTV \\\midrule 
T1w& $29.89 \pm 9.35$ & $ 30.51 \pm 10.76$ & $30.71 \pm 10.84$& 
$\bm{31.19 \pm 9.41}$
\\ 
T2w & $28.10 \pm 8.02$ & $28.71 \pm 8.85$& $28.98 \pm 8.96$& 
$\bm{29.58 \pm 7.86}$
\\ 
PDw & $28.37 \pm 9.85$ & $28.59 \pm 10.02$& $28.95 \pm 9.85$& 
$\bm{29.96 \pm 8.93}$
\\\bottomrule
\end{widetable}
\label{tab:superres:psnr}
\end{table*}

\begin{table*}
\fontsize{8}{7.2}\selectfont
\centering
\caption[Results for segmenting super-resolved images]{Results for
segmenting super-resolved images (for 576 subjects). Dice scores for
different reconstruction methods were computed for GM, WM and CSF tissue
segmentations, using HR segmentations as references. Results shown as
mean$\pm$sd.}
\begin{widetable}{\textwidth}{*{5}{c}} \toprule
Tissue class & BS & FOT & TV & MTV \\\midrule 
GM& $0.831 \pm 0.001$ & $ 0.838 \pm 0.001$& $0.842 \pm 0.002$& 
$\bm{0.887 \pm 0.002}$
\\ 
WM & $0.866 \pm 0.001$ & $0.874 \pm 0.002$& $0.878 \pm 0.001$& 
$\bm{0.891 \pm 0.001}$
\\ 
CSF & $0.842 \pm 0.002$ & $0.852 \pm 0.001$& $0.856 \pm 0.001$& 
$\bm{0.885 \pm 0.002}$
\\\bottomrule
\end{widetable}
\label{tab:superres:ds}
\end{table*}

\end{description}

\begin{figure*}
\centering 
\includegraphics[width=0.95\textwidth]{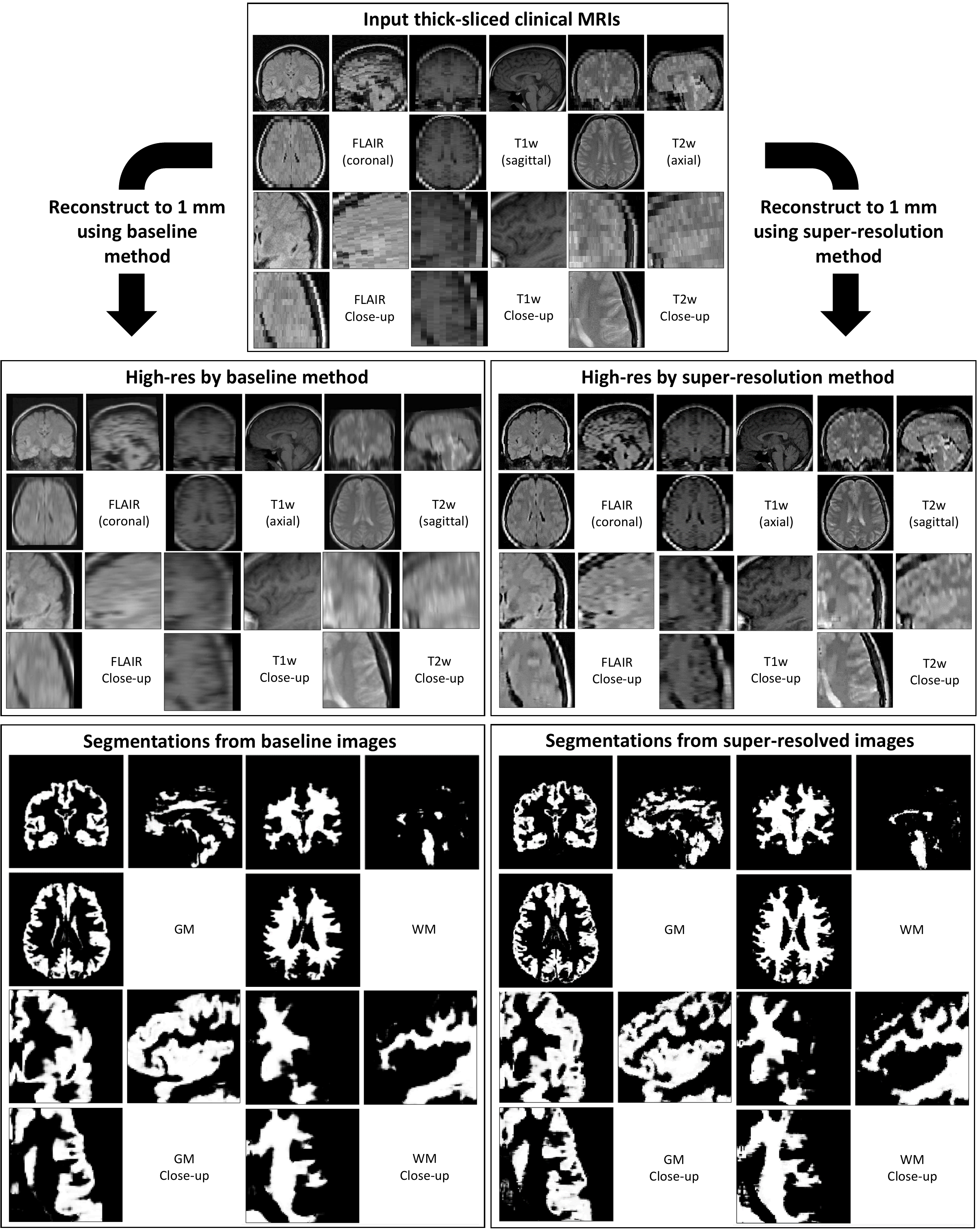}
\caption[Example results for baseline vs super-resolution]{Example
results for baseline vs super-resolution, for a randomly selected
patient. The box on top shows the three input clinical scans (FLAIR, T1w
and T2w). These three scans are reconstructed to 1 mm isotropic voxel
size, using either the baseline (left) or the super-resolution method
(right). It is clear that the multi-channel segmentation output (GM and
WM), produced from the super-resolved images have better anatomical
detail. For example, the WM has a clearer delineation close to the
cortex.}
\label{fig:superres:ex-hospital}
\end{figure*}

\begin{figure*}
\centering 
\includegraphics[width=\textwidth]{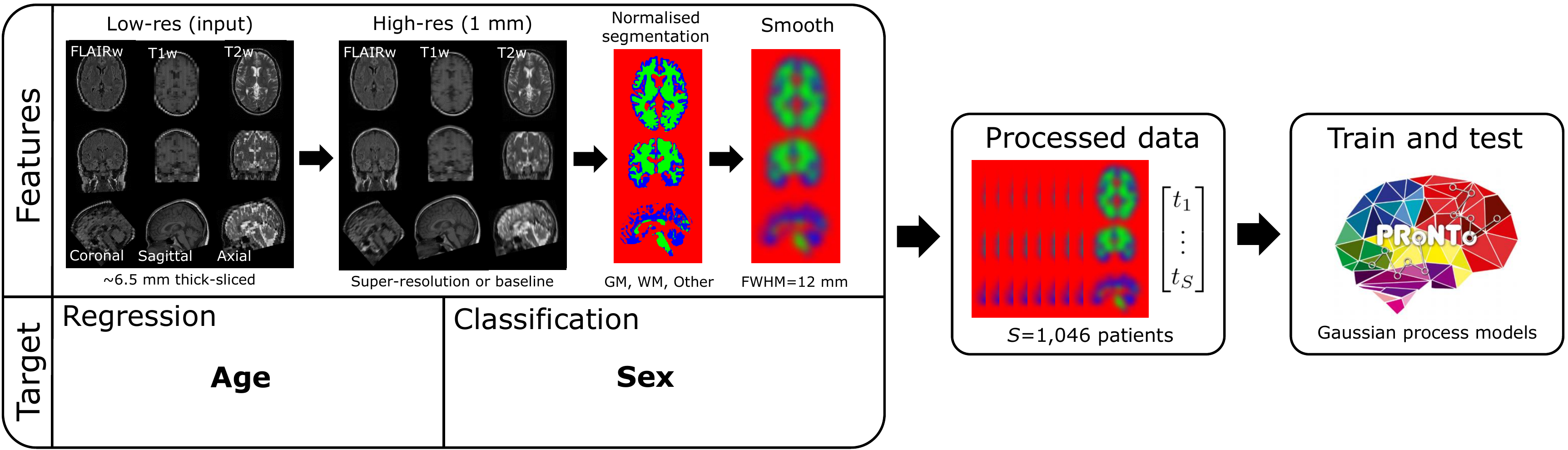}
\caption[Evaluation of the super-resolution method on clinical
data]{Evaluation of the super-resolution method on clinical data.
Feature vectors are obtained by concatenating smoothed, gray matter
(GM), white matter (WM) and other segmentations. These segmentations are
produced from multi-channel MRIs that has been reconstructed to 1 mm
isotropic voxel size using either super-resolution or a baseline method.
The prediction targets are either the age or sex of the patients.
Predictive performance is then evaluated, for both, methods using
Gaussian process models in the PRoNTo toolbox.}
\label{fig:superres:agesexprediction}
\end{figure*}


\subsection{Clinical Data}

In the previous section we showed that multi-channel super-resolution
outperforms some established single-channel techniques, on simulated
data. This result is promising as we now move on to evaluating the
method on real, clinical-grade MR data -- a far more challenging
scenario. Conversely to simulated data, there is no ground truth for
clinical data. Therefore, we propose to evaluate implicitly the quality
of the reconstructed images by using them as input to machine learning
models to predict known, noise-free features: age and sex.

\begin{figure*}[t]
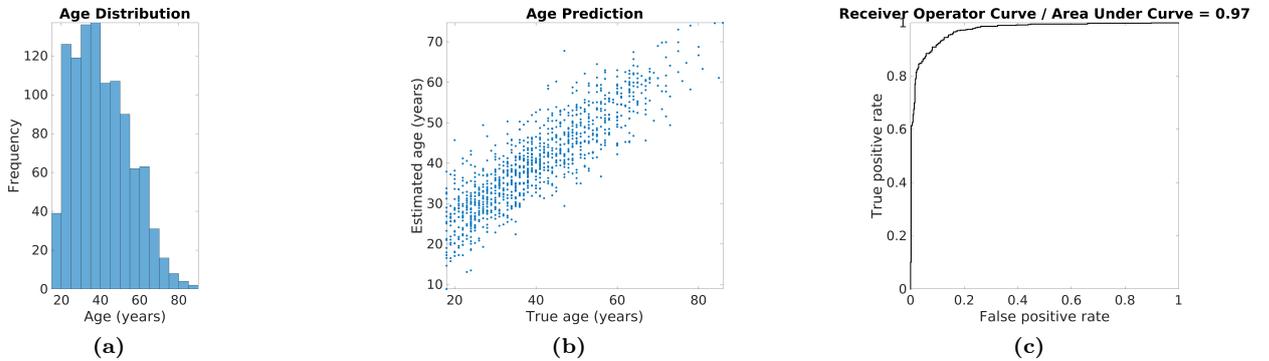

\centering
\subfloat[]{\includegraphics[width=0.32\textwidth]
{age-dist}\label{fig:superres:hosp-distage}} \hfil
\subfloat[]{\includegraphics[width=0.32\textwidth]
{age-pred}\label{fig:superres:age-pred}} \hfil
\subfloat[]{\includegraphics[width=0.32\textwidth]
{sex-pred.pdf}\label{fig:superres:sex-pred}}
\caption[Plots of results for evaluating the super-resolution
model]{Results for evaluating the super-resolution model. (a) Age
distribution of the 1,046 patients in the clinical dataset. Mean value
is $40.1$ and standard deviation is $14.3$ years. (b) Scatter plot of
the individual predictions for the regressions task. (c) ROC curve for
the classification task, where an AUC of 0.97 was obtained.}
\end{figure*}

The dataset we used consists of 1,046 patient's MRIs, with 615 males and
431 females. The dataset was acquired on a diversity of scanners and
clinical indications at UCLH (University College London Hospitals,
London, UK) in the context of routine clinical care. The dataset
comprises three contrasts (T1w, T2w and FLAIR), each with a different
thick-slice direction; the average slice-thickness over the whole
dataset is 6.5 mm. Commonly, the images have only partial brain coverage
in the thick-slice direction, with the outer most slices excluded. The
age distribution of the dataset is shown in Figure
\ref{fig:superres:hosp-distage}.

The gist of our validation relies on training machine learning models to
predict age and sex from tissue segmentations of patient MR images. It
is now well known that these features can be very accurately predicted
from brain images
\citep{monte2018comparison,smith2019estimation,he2018deep,cole2017predicting}.
 Here, we follow the procedure of \cite{monte2018comparison}. Since 
(accurate) normalised segmentations capture relevant and important 
anatomical features of the MRIs, predictive accuracy can be used as a 
measure of the quality of a super-resolution model.

For each subject in the dataset, the LR images were super-resolved to 1
millimetre isotropic with BS (4th order) or MTV, and segmented using
SPM12's unified segmentation routine \citep{ashburner2005unified}, with
default parameters. This routine outputs normalised, non-modulated, GM,
WM and Other (1 - GM - WM) maps, that were smoothed with a Gaussian
kernel of 12 mm full-width at half-maximum (FWHM). The concatenated
smoothed maps were used as a feature vector for machine-learning. Two
Gaussian process models were trained, using 10-fold cross-validation, to
predict age and sex from this feature vector. The PRoNTo toolbox
\citep{schrouff2013pronto} was used to make these predictions. Using a
convolutional neural network based model could have been an option, but
recent studies suggest that kernel methods performs comparably for
predictive tasks similar to the ones performed here \citep{he2018deep}.

The results of both the regression and classification task are shown in
Table \ref{tab:superres:pred}. Age regression results are reported in
years using the root mean square error (RMSE), error standard deviation
(SD), mean error (bias) and Pearson's correlation coefficient; sex
classification accuracy is reported in percentage, and the lower and
upper bound over the 10 folds are included. Furthermore, for the MTV
case, a scatter plot of the individual predictions for the regressions
task is shown in Figure \ref{fig:superres:age-pred}, and a receiver
operator curve (ROC) for the classification task is shown in Figure
\ref{fig:superres:sex-pred}. Images and non-smoothed native space
segmentations, for an example patient, are shown in Figure
\ref{fig:superres:ex-hospital}. Our results make it quite clear that the
segmentations produced from the super-resolved images have more
predictive power. This is true for both the regression and
classification task (\emph{c.f.} Table \ref{tab:superres:pred}). In
particular for the classification task, where the improvement in
accuracy is 3.3 percentage points. It also evident from the example
segmentations in Figure \ref{fig:superres:ex-hospital} that anatomical
features are more clearly defined in the super-resolved segmentations.
Furthermore, comparing, \emph{e.g.}, the super-resolved close-up of the
T1w image with its baseline counterpart, it can be seen that the
mismatched fields of views have been filled in for the super-resolution
case. Finally, the runtime of the algorithm to super-resolve three LR
images, as was performed in this evaluation, is under 30 minutes on a
modern workstation. The runtime scales linearly with the number of
channels.

\begin{table*}[t]
\fontsize{8}{7.2}\selectfont
\centering
\caption[Results for predicting age and sex from normalised brain
segmentations]{Results for predicting age and sex from normalised brain
segmentations (for 1,046 patients). We compared the proposed
super-resolution model to a baseline method, which reslices using 4th
order b-spline interpolation. Age regression results are reported in
years using the root mean square error (RMSE), error standard deviation
(SD), mean error (bias) and Pearson's correlation coefficient. Sex
classification accuracy is reported in percentage, where the lower and
upper bound over the 10 folds are included.}
\begin{widetable}{\textwidth}{c | c c c c | c c c} \toprule
& \multicolumn{4}{c|}{Age (years)} & \multicolumn{3}{c}{Sex (\%)} \\
Method & RMSE & SD & Bias & Correlation & Accuracy & Lower bound & 
Upper 
bound 
\\\midrule
Baseline         & 6.91 & 9.89 & -1.73 & 0.85 & 87.8 & 85.6 & 89.6 \\
Super-resolution & 6.32 & 8.66 & -1.18 & 0.88 & 91.1 & 89.2 & 92.7
\\\bottomrule
\end{widetable}
\label{tab:superres:pred}
\end{table*}

\section{Discussion}

This paper presents a super-resolution tool that can be applied to large
datasets of clinical MR scans. Commonly in such datasets, each patient
has a collection of images acquired with different MR sequences.
Currently, when performing some multi-channel analysis, these images are
often simply interpolated to the same size. The model proposed here is
an alternative to interpolation that better leverages these multiple MR
contrasts. The model builds on a principled probabilistic generative
model. The novelty of the model lies in the MTV prior, which allows a
joint probability distribution across MR contrasts to be modelled. All
model parameters are set automatically, and no fine-tuning is necessary,
allowing large datasets to be processed with variable slice-thicknesses
and MR contrasts. We showed that, when the super-resolution model is
used as a preprocessing step, subsequent machine-learning tasks have
improved results.

Data-driven models currently show state-of-the art performances in many
image processing tasks and could be an alternative to the approach
proposed in this paper. However, generalisability is still an issue with
many such methods, as they excel in scenarios where the unseen data is
close to data the model was trained on but does not extrapolate well on
out-of-sample test data. This is because data-driven models construct an
empirical prior from the data, which leads to a flexible, highly
parametrised model -- where minor prior assumptions are necessary -- but
where overfitting to a training population can be an issue. This is 
especially risky with pathological imaging, because pathology adds 
further diversity on already hugely diverse normal biology. Model-based
methods, on the other hand, require prior assumptions to be made. To
design priors as flexible as ones learnt by data-driven techniques is
extremely challenging, and simplified assumptions are therefore often
made. If the prior assumption is close to the data generating process,
and the forward model and statistical properties of the data is too,
then model-based methods can perform well on out-of-sample data (and
will not require any retraining). Interestingly, methods have been
developed that combine model- and data-driven approaches
\citep{adler2018learned,dalca2019unsupervised,brudfors2019nonlinear}.
This combination would be an interesting future direction for the
super-resolution model proposed in this paper.

The model we have proposed could be of value in translating methods that
have shown good results on research data to clinical imaging. For
example, many techniques based on machine (deep) learning show promising
results on analysing neuroimaging data \citep{litjens2017survey}.
However, a model trained on HR data may struggle when given as input LR
clinical data. Applying our super-resolution model as a preprocessing
step, reconstructing HR versions from the LR input, could facilitate
this transition. The evaluation also showed that the tissue segmentation
performance of a widely used neuroimaging package improves when the
multi-channel input images are super-resolved, compared with simply
being interpolated. Furthermore, the model could easily be used for
multi-channel denoising by replacing the projection matrix in
\eqref{eq:superres:cond} with identity\footnote{Running the denoising
version of the model is an option in our software implementation.}.

However, metric scores such as those used in this study are specific to
the data that was used to evaluate the model. Any claim that a method
generalises to the huge variability present in clinical MR scans should
therefore be taken with a pinch of salt. This is not only because the
imaging data varies greatly: in image contrast, intra-subject alignment
and voxel size; but also because the scanner acquiring the images may
not have computed the correct voxel sizes, slice-thicknesses,
\emph{etc}. In our evaluation, we used a clinical dataset with a large
variability among patient scans. Still, this is no guarantee that our
model will work on any clinical MR data. However, data specificity is
likely to be greater the more flexible the model is, such that highly
parameterised models may suffer more from this specificity.

The model proposed in this paper could be made, possibly, more robust in
a few ways. One is related to our assumptions regarding the slice
profile and slice gap. These parameters are highly variable and assuming
them as fixed, as is currently done, can lead to inexact super-resolved
images. Of course, a user could change these parameters themselves if
they know their values. However, they sometimes cannot be obtained. A
principled solution to this problem would be to extend the
parametrisation of the projection matrices to model both slice gap and
profile. Although a non-trivial modelling problem, this would allow us
to estimate the most likely such parameters. Misalignment between scans
could be another explanation for poor super-resolved images, as our
edge-based prior distribution is highly dependent on well registered LR
images. Rather than performing an initial rigid registration of the LR
images (as is currently done) improved alignment could be achieved by
modifying the forward model in \eqref{eq:superres:noisemodel} to
incorporate a rigid transformation. The optimal parameters could then be
found by Gauss-Newton optimisation, similar to what is done in
\citep{ashburner2013symmetric}.

A well known fact is that TV introduces stair-casing effects on flat
areas, which are abundant in brain MRI. However, the MTV regularisation
proposed in this paper seems not to suffer from such artefacts. This is
probably because MTV uses gradients distributed over all contrasts,
which often have orthogonal thick-slice directions. Hence, an image area
containing flat gradients in one channel may very well have more
informative gradients in another channel. Another factor that suppresses
these stair-casing artefacts is our parametrisation of the differential
operator, which computes the gradient using both the forward and the
backward finite differences.

Many different algorithms are available to solve nonsmooth optimisation
problems such as the one that arises when using a nonsmooth TV prior.
Here, we used an ADMM algorithm because it is relatively fast and easy
to implement, and since the quadratic problem that arises can be easily
solved with a multigrid solver that assumes a stationary penalty over
gradients. However, it makes proper marginalisation of the (latent) HR
image difficult. An alternative could be to use an approach based on
reweighted least-squares (RLS)
\citep{daubechies2010iteratively,bach2012optimization,grohs2014tv},
which makes use of the bound:
\begin{align} 
\norm{\vt{z}}_2 = \min_{w > 0} \left\{ \frac{1}{2w} \norm{\vt{z}}_2^2 +
\frac{w}{2} \right\}.
\end{align} 
We could substitute this bound, with $\vt{z}=\vt{D}_n\vt{Y}$, in the
joint log-distribution. However, this would require extending the
multigrid solver to handle non-stationary penalties. Furthermore,
instead of looking for a MAP solution, an approximate Gaussian posterior
that factorises over voxels and channels could be obtained using
variational Bayes \citep{attias2000variational}. When used in
combination with RLS, this posterior would be bounded by a Gaussian.
Finally, commonalities between channels do not solely reduce to edges;
intensities co-vary as well. Mutual information between channels could
be taken into account by introducing an additional, independent, prior
over the reconstructed image in the form of a multivariate Gaussian
mixture \citep{2019arXiv190805926B}. This would be, in practice, a joint
reconstruction and segmentation framework. When multiple images are
available, commonalities across subjects could be learnt as well by
using a learnable shape and intensity model
\citep{blaiotta2018generative,ashburner2019algorithm}. Such an extended
Bayesian framework may eventually be able to compete with (deep)
learning-based approaches, or could even be defined as to include such
models in the generative process \citep{brudfors2019nonlinear}.

\subsubsection*{Clinical Data:} The clinical data that was used in
this paper is a sample of anonymized MR studies obtained within the
clinical routine, for which ethical and regulatory approval has been
obtained from the Health Research Authority (HRA)

\subsubsection*{Acknowledgements:} MB was funded by the EPSRC-funded UCL
Centre for Doctoral Training in Medical Imaging (EP/L016478/1) and the
Department of Health’s NIHR-funded Biomedical Research Centre at
University College London Hospitals. YB was funded by the MRC and Spinal
Research Charity through the ERA-NET Neuron joint call (MR/R000050/1).
PN was funded by the Wellcome Trust and the NIHR UCLH Biomedical
Research Centre. MB and JA were funded by the EU Human Brain Project's
Grant Agreement No 785907 (SGA2).

\section*{References} 
\bibliographystyle{elsarticle-harv} 
\bibliography{bibliography}

\end{document}

%% file: front.tex
\begin{frontmatter}

\journal{NeuroImage}

\title{A Tool for Super-Resolving Multimodal Clinical MRI}

\author[WCHN]{Mikael Brudfors\corref{cor}}
\author[WCHN]{Ya\"{e}l~Balbastre}
\author[ION]{Parashkev Nachev}
\author[WCHN]{John Ashburner}

\address[WCHN]{Wellcome Centre for Human Neuroimaging, University 
College London, London WC1N 3BG, UK}
\address[ION]{Institute of Neurology, University College London, 
London WC1N 3BG, UK}

\cortext[cor]{Corresponding author: Mikael Brudfors \\ 
\texttt{\href{mailto:mikael.brudfors.15@ucl.ac.uk}{mikael.brudfors.15@ucl.ac.uk}}
}

\begin{abstract} 
We present a tool for resolution recovery in multimodal clinical
magnetic resonance imaging (MRI). Such images exhibit great variability,
both biological and instrumental. This variability makes automated
processing with neuroimaging analysis software very challenging. %
This leaves intelligence extractable only from large-scale analyses of
clinical data untapped, and impedes the introduction of automated
predictive systems in clinical care. The tool presented in this paper
enables such processing, via inference in a generative model of
thick-sliced, multi-contrast MR scans. All model parameters are
estimated from the observed data, without the need for manual tuning.
The model-driven nature of the approach means that no type of training
is needed for applicability to the diversity of MR contrasts present in
a clinical context. We show on simulated data that the proposed approach
outperforms conventional model-based techniques, and on a large hospital
dataset of multimodal MRIs that the tool can successfully super-resolve
very thick-sliced images. The implementation is available from
\url{https://github.com/brudfors/spm_superres}.
\end{abstract}

\begin{keyword}
Generative modelling, Super-resolution, MRI, Clinical data, 
Multi-channel total variation
\end{keyword}

\end{frontmatter}